%% file: ms.tex
\documentclass[12pt]{article}
\usepackage{aaspp4}
\usepackage{epsfig}
\usepackage{astrobib}
\input nkstddefs

\input href_unlinked

\begin{document}

\title{A Photometric Study of the Supercluster MS0302 with the UH8K CCD Camera:
Image Processing and Object Catalogs}
\author{N.~Kaiser, G.~Wilson, G.~Luppino, H.~Dahle}
\affil{Institute for Astronomy, University of Hawaii}
\authoraddr{Institute for Astronomy, University of Hawaii, 2680 Woodlawn Drive, Honolulu, Hawaii 96822}

\input abstract

\input introduction.tex

\input prelim.tex
\input registration.tex

\input extinction.tex

\input psfmodel.tex

\input warping.tex

\input noise.tex

\input objectdetection.tex

\input photometry.tex
\input conclusions.tex



\input references.tex

\appendix
\input acfregister

\input interpolation

\input lcformat

\end{document}

%% file: nkstddefs.tex
\def \beginfig          {\begin{figure}}
\def \endfig            {\end{figure}}

\def \begineq           {\begin{equation}}
\def \endeq             {\end{equation}}

\def \d {\partial}

\def\gtorder{\mathrel{\raise.3ex\hbox{$>$}\mkern-14mu
             \lower0.6ex\hbox{$\sim$}}}
\def\ltorder{\mathrel{\raise.3ex\hbox{$<$}\mkern-14mu
             \lower0.6ex\hbox{$\sim$}}}

\def \lsim {\ltorder}

\def \hide#1{}
\def \half {{1\over 2}}



%% file: href_unlinked.tex
\newcommand{\astrophref}[1]{{\tt http://xxx.lanl.gov/abs/astro-ph/#1}}

\newcommand{\href}[2]{#1}

%% file: abstract.tex
\begin{abstract}
We describe in detail the processing of a set of
images of the $z = 0.42$ supercluster MS0302 taken with the UH8K
camera at CFHT. The result of this is a pair of seamless combined V-
and I-band images of the field, along with a characterization of the
noise properties and of the point spread function (PSF), and catalogs
of $\simeq 30,000$ faint galaxies.  The analysis involves the following
steps: image preparation; detection of stars and registration to find
the transformation from detector to sky coordinates; correction for
extinction and/or gain variations; modeling of the PSF; generation of
images with a circular PSF; image warping and averaging; modeling of
the noise auto-correlation function; faint object detection, aperture
photometry, and shape measurement.  The shear analysis is described
elsewhere.  
\end{abstract}

%% file: introduction.tex
\section{Introduction}
\label{sec:introduction}

The data described in the paper were taken as part of a program to
measure weak gravitational lensing.  
Reducing these data, taken with the UH8K mosaic camera, has proved to be
a complex process.  
However, after a good deal of experimentation, we
feel we have in place a fairly reliable, accurate and largely automated 
procedure.
The purpose of this paper is to describe in detail how the
summed images and catalogs of faint objects were constructed,
with the weak lensing analysis being described elsewhere \cite{kwl+98}.
This paper also serves as a technical reference for our group's
UH8K `blank field' survey, 
which comprises 6  similarly deep fields \cite{wkl99}, and
which has been analyzed in much the same way. 
The paper may also possibly be useful for others attempting to reduce data from 
the UH8K or other similar mosaic cameras, and to this end we highlight
some of the inadequacies of our current approach.

The target field, centered roughly on 
$\rm{RA} = 3^h 5^m 24^s.0, \rm{DEC} = 17^\circ 18' 0''.0$, (J2000)
contains three prominent clusters in
a supercluster at $z \simeq 0.42$.
CL0303+1706, 
was detected optically by \citeN{dg92}, and an Einstein IPC pointed observation
revealed the
presence of the two neighboring clusters
MS0302+1659 and MS0302+1717 with redshifts $z= 0.426, 0.425$
(\citeNP{emss90}; citeNP{emss91}). 
All three clusters have numerous measured redshifts and form a physically
associated complex at $z \simeq 0.4$, which fits snugly within the
$0^\circ .5$ square field.

The data were taken at the 3.6m CFHT on the nights 22-24 September, 1995.
A total of 17 exposures (11 in I-band, 6 in V-band), each of 900 seconds  
integration time were obtained.  In addition, a number of dark frames and
dome flats were taken.  
Standard stars observations were interspersed between the science
exposures,
which were performed with small semi-random offsets from
the nominal field center (typically $40''$ offsets) to 
allow the
removal of cosmetic defects and the generation of a 
seamless contiguous combined image.  The image quality was excellent, with
stellar FWHM $\simeq 0''.6$.

The layout of the paper is as follows:
\S\ref{sec:preliminaryreduction}: preliminary reduction; 
\S\ref{sec:registration}: detection of stars and registration to find the transformation
from detector to sky coordinates; 
\S\ref{sec:extinction}: correction for extinction and/or gain variations;
\S\ref{sec:psfmodeling}: modeling of the point spread function (PSF) and
generation of images with a circular PSF;
\S\ref{sec:warping}: image warping and averaging;
\S\ref{sec:noiseproperties}: modeling of the noise auto-correlation function (ACF);
\S\ref{sec:objectdetection}: faint object detection;
\S\ref{sec:photometry}: aperture photometry and shape measurement.

%% file: prelim.tex
\section{Preliminary Reduction}
\label{sec:preliminaryreduction}

The preliminary reduction of the data was fairly straightforward.
We first subtracted from each image a bias, this being a linear ramp fit to
the pixel values in the over-scan region. Then
for each chip we computed a median of several dark frames and subtracted
this from each science image. Then,
for each passband and for each chip we 
computed a median `super-flat' from all the images obtained with that chip, 
and divided each of the science images by this
sky flat.   Note that this will have reduced the sensitivity of these
images to very extended diffuse flux.

A number of bad columns and other cosmetic defects were clearly visible as 
high contrast features in the median sky flats.  A simple algorithm
was used to identify these abnormal pixels and the corresponding
pixels in the science images were flagged as unreliable (this is done
by setting the pixel value to the `magic' value of -32678 (the most negative
number expressible in the 16 bit signed integer pixel format we have used);
our image processing software recognizes this value and, generally speaking,
the result of any computation involving a magic input value is also
set to be magic (the major exception to this rule being the image
co-addition).

The science images were then all visually inspected, and a number of
further cosmetic defects were flagged as bad data.  
The data in one of the chips
--- that in the NW corner --- was found to be seriously compromised
and these data were discarded.
 
The images were then cropped from their original, slightly oversize, format 
to 2048 by 4096 format.  The images in
the top row (which are read out in the opposite direction to those in
the bottom row) were then inverted so that North was approximately aligned
with the $+y$ direction (i.e.~the slow readout direction) on all the images.

Finally, we subtracted a smoothed local sky estimate
determined from the heights of minima of the images.  More
specifically, if the locations of the minima are $x_i$
and have values $f_i$ then we generated the pair of
images $n(x) = \sum \delta(x - x_i)$ and 
$f(x) = \sum f_i \delta(x - x_i)$, convolved both with a 
32 pixel Gaussian to make the smoothed images $f_s$ and $n_s$ and
then subtracted $f_s / n_s$.
As with the `super-flat'
generation, this will inevitably have suppressed long wavelength
features such as possible highly extended diffuse emission from the clusters.

%% file: registration.tex
\section{Astrometric Registration}
\label{sec:registration}

We now describe how we solved for the mapping from pixel or
`detector' coordinates onto 
a planar projection of the sky.

In the UH8K camera the chips are laid out in two rows each of
four 2K by 4K chips laid side by side.  There are gaps of about
60 pixels between the sides of the chips and 
about 20 pixels between the two rows.
However, the chips are not precisely laid out on the grid but
are slightly rotated and shifted with respect to an ideal tiling grid.  
In addition to
the somewhat irregular chip layout, the CFHT suffers from a field distortion
introduced by the telescope wide-field corrector; a pin-cushion
distortion with shift amplitude of about 40 pixels, giving
a radial shear of $\gamma \simeq 7.2 \times 10^{-3}$,
at the corners of the roughly 1/2 degree square field. 

The registration procedure will impact the weak shear measurement
in two ways: First, the telescope distortion, if uncorrected, will mimic the
effect of a (negative mass) gravitational lens.  This effect is 
relatively easy to deal with.  Secondly, and potentially
much more damaging, is the gross anisotropy of the summed image PSF that
can result from errors in the
registration.  Initially we tried to model the distortion
assuming a `rigid detector' model, with fixed parameters describing the
layout of the rectangular chips in detector plane,
but found that this did not yield adequate precision.
To obtain sufficient accuracy we found we needed to
relax the assumptions of the model; instead we assumed that for
each 2K$\times$4K image there is
some a low-order polynomial
mapping from pixel coordinates to the rectilinear sky coordinates,
but we do not assume that there is any static relationship between
the mapping from exposure to exposure --- i.e~we are assuming that the
telescope/detector system
can deform in a smooth but otherwise fairly 
arbitrary manner between exposures.
This  `Jello detector' model allows for
both the telescope distortion and chip layout as well as
image deformations associated with filters, atmospheric refraction,
thermal expansion and mechanical strain.

\subsection{Astrometric Reference System}

We solved for the parameters of these polynomial mappings --- the `astrometric solution'
--- by minimizing the residuals in the predicted sky positions of
a set of reference stars. In principle, given certain conditions on the
geometry of the telescope pointings, one should be able to perform 
an `internal' solution using only the CFHT images and without 
reference to any external astrometric system 
(one should then be able to tie this into
any other chosen absolute astrometric coordinate system by inspection 
of the coordinates of any pair of stars).  With sufficiently
large offsets between exposures, this works quite well, but
with the rather small offsets between
exposures employed here unfortunately, 
this turns out to be quite unstable; it is 
fairly easy to find solutions which map the star positions onto each other
to great accuracy, but the solutions tend to have unacceptably large
large-scale field distortions.  What we did instead was to incorporate
the external astrometric information from the start.

The external astrometric data base to which we have chosen to refer our coordinate
system is the USNOA catalog \cite{usnoa}.  Unfortunately 
there are rather few
stars which are both present in the USNOA catalog and are non-saturated in the
CFH images.  
This seems to be because to be included in the USNOA catalog
an object had to be detected in both red and blue passbands, so many
stars which were actually detected in the red
drop out of the final catalog.  It is also a consequence of the 
rather long exposure
times forced upon us owing to the rather long readout time for the
UH8K mosaic; with detectors like the CFH12K the readout time is
greatly reduced and this is much less of a problem.
To work around
this problem we extracted the digitized sky survey 
\cite{dsswebsite} red image covering
our field and ran our own object detection algorithm which detects most of the 
USNOA objects, which are predominantly stars, 
and which were originally derived from the same Palomar plates,
as well as a substantial number of stars which did not survive the
USNOA selection criteria. In the final astrometric solution we
used only fairly bright, but non-saturated, stars from the CFH
images, and the overlap with
the augmented reference catalog is such that we typically found about
50 stars in common per
$2$K$\times 4$K CFH image. 

The digitized sky survey FITS image 
contains a `world-coordinate system' definition in the
header which relates DSS pixel indices to celestial coordinates \cite{gc95}. 
After
choosing the nominal field center $(\alpha_0, \delta_0)$ = (3:5:24.0, 17:18:0.0)
we then generated the orthographic sky coordinates $r$ (see below) 
for each of the DSS stars.
Comparing these with the USNOA catalog stars we noticed a small and slowly
varying systematic discrepancy between the derived sky coordinates (with amplitude
on the order of $0''.25$).  We modeled this as a low order
polynomial correction
\begin{equation}
\label{eq:usnoatodssmapping}
r_{\rm USNOA} = r_{\rm DSS} + 
\sum\limits_{l=0}^{l_{\rm max}} 
\sum\limits_{m=0}^{l} 
a_{lm} f_{lm}(r_{\rm DSS})
\end{equation}
where the mode functions are given by
\begin{equation}
\label{eq:flmdefinition}
f_{lm}(r) = r_0^{l-m} r_1^m 
\end{equation}
so e.g.~for $l=2$, and writing $r_0 = x$, $r_1 = y$,
the modes are the 3 quadratic functions $x^2$, $xy$, $y^2$;
for $l=3$, the modes are the 4 cubic functions $x^3$, $x^2y$, $xy^2$, $y^3$
and so on.
We then applied this correction to the DSS
star coordinates, with $l_{\rm max} = 3$, 
to bring them into agreement with the USNOA system.

\begin{figure}[htbp!]
\centering\epsfig{file=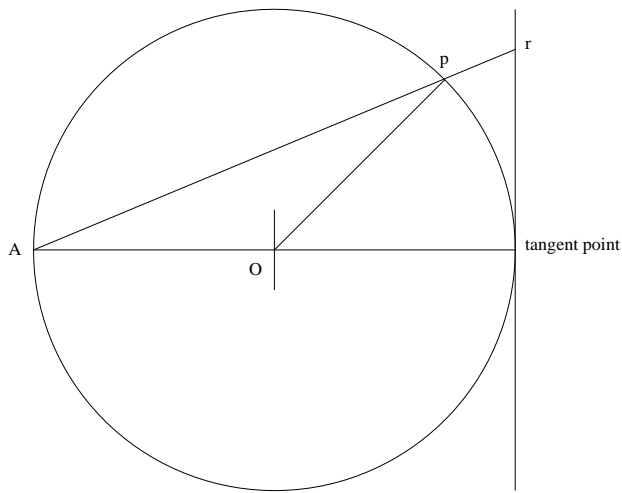,width=0.5 \linewidth}
\caption[Orthographic Projection]{Illustration of the orthographic projection we have adopted
where we have rotated the coordinate system so that the
field center $(\alpha_0, \delta_0)$, 
or tangent point, lies along the $x$-axis.  A point $p = (\alpha, \delta)$
on the sky (represented here by the unit sphere) is projected onto the
orthographic coordinate plane by projecting the line from the
antipode of the tangent point through $p$ into the tangent plane at
point $r$. To fully define the $r$ coordinate system it is
necessary to specify the celestial coordinates of the field center
$(\alpha_0,\delta_0)$, and a rotation angle giving the orientation of the
axes of $r$-coordinate system relative to the latitude and longitude
directed unit vectors at the field center, and a scale factor.}
\label{fig:projection}
\end{figure}

The projection we adopted is the `orthographic' projection
which is that particular
stereographic projection of the sky illustrated in figure
\ref{fig:projection}.
This projection is shape (but not size) preserving, which
is convenient for weak lensing studies; though had we chosen any of the
other standard projections the induced shape distortions would have
been on the order of $\theta^2 \sim 10^{-5}$
and would be negligibly small for the field size here,
and similarly, the scale change across the field for the projection we have
adopted is on the same order. 

To fully define the orthographic projection it is necessary to specify
not just the tangent point, but also a rotation angle giving the
orientation of the tangent plane.  Specifically, this is the rotation
of the Cartesian axes $r_0, r_1$ relative to the longitude and
latitude directed unit vectors at the tangent point; this defines the
pixel coordinate axes in the final summed images.  
A natural choice would be
to set this so that the `$y$'- or $r_1$-axis is aligned with a line
of longitude, so that, at the field center,
North is `up'.  However, the camera axis was slightly
misaligned with North, so with this choice of sky coordinates
the bleeding of saturated stars along the slow chip axis would
then be slightly tilted with respect to our final coordinate axes
which would be awkward later when we come to mask out these
features. To avoid this, we adopted a rotation of $2^\circ .635662$ to
approximately align the star trails with the $r_1$ axis.

\subsection{Astrometric Solution}

To solve for the image mapping parameters with the MS0302 data we used bright (but
non-saturated) stars.  Stars brighter than about 20th magnitude are well separated from galaxies 
in the size-magnitude plane.
We typically extracted 100 or so stellar objects per image, or on the order of
1000 stars for a complete mosaic, and of these about one half were
reliably measured in the astrometric reference catalog.
Stellar centroids were measured to fractional pixel precision using the
interpolation scheme described in appendix \ref{sec:2dinterpolation}.
We solved simultaneously for a set of low order spatial
polynomials, one per $2K \times 4K$ image, which map pixel coordinates
onto the sky.  The solution was obtained by a sequence of refined
least squares minimizations.
The relatively accurate CCD mosaic star positions
ensure that the CFHT images map onto each other with high
precision, while the external catalog
serves to damp down the kinds of artificial distortion one
would otherwise obtain with a purely internal solution. 

\begin{figure}[htbp!]
\centering{\epsfig{file=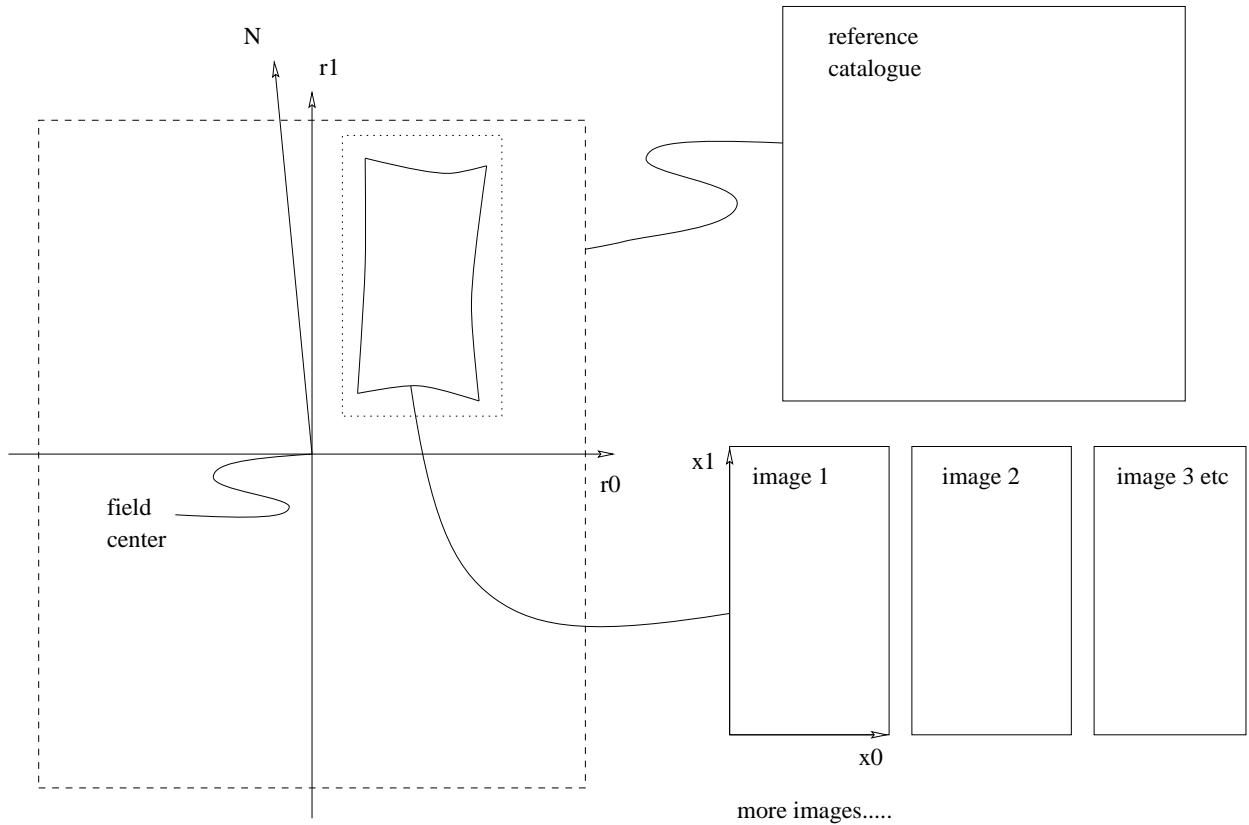,width=\linewidth}}
\caption[Image Mapping]{Illustration of the mapping from CFHT image pixel coordinates
and reference catalog coordinates onto the sky. 
The orthographic coordinates from the reference catalog 
map directly onto $r$-space with no distortion (but with considerable
measurement error).  Each of the CFHT science images maps onto
the plane in a distorted manner, shown grossly exaggerated in the
figure, and which we describe by a low-order
spatial polynomial as described in the text. Also shown dotted is the
slightly oversize bounding boxes that we compute; these are used
when we come to apply the warping to the images to identify which
images contribute to a given patch of the sky. 
}
\label{fig:geometry}
\end{figure}

Specifically, we modeled the mapping from pixel coordinates $x_{pi}$ 
(this being the position
of the $p$'th star on the $i$'th image) to $r = r_{\rm USNOA}$ coordinates
as a cubic polynomial
just as in (\ref{eq:usnoatodssmapping}) with
\begin{equation}
\label{eq:mapping}
r_p = x_{pi} + \sum\limits_{l=0}^{l=3} 
\sum\limits_{m=0}^{m=l} a_{ilm} f_{lm}(x_{pi}) + e_{pi}
\end{equation}
with mode functions as in (\ref{eq:flmdefinition}) and
where $e_{pi}$ is the observational error, which we assume to have
an isotropic 2-D Gaussian PDF with scale length $\sigma_{pi}$, whereas
for the reference catalog (the DSS
catalog, corrected as described above, and to which we ascribe the 
index $i=0$) 
\begin{equation}
r_p = x_{p0} + e_{p0}
\end{equation}
or equivalently we can say that (\ref{eq:mapping}) applies for all $i$ with
the understanding that $a_{0lm} = 0$.
This mapping is illustrated in figure \ref{fig:geometry}.

The astrometric solution is that set of star positions 
$r_p$ and parameters $a_{ilm}$ which minimize the
sum of the squared residuals:
\begin{equation}
\chi^2 = \sum_p\sum_i e_{pi}^2 / \sigma_{pi}^2 =
\sum_p\sum_i (r_p - x_{pi} - \sum\limits_{l,m} a_{ilm} f_{lm}(x_{pi}))^2 / \sigma_{pi}^2
\end{equation}
which is quadratic in the parameters $a_{ilm}$, $r_p$.
For our pointings 
this yields the well conditioned set of linear equations
\begin{equation}
\label{eq:leastsquares}
\partial \chi^2 / \partial a_{ilm} = 0; 
\quad \quad
\partial \chi^2 / \partial r_p = 0
\end{equation}
which we solved by LU decomposition for the mode amplitudes $a_{ilm}$ and star
positions $r_p$.
It is important here to allow for the
the fact that the DSS coordinates
are relatively imprecise: $\sigma_0 \sim 0''.25$ 
as compared with rms precision for the star centroid coordinates on the
CFH images of about $0''.007$ (see below).  Empirically, we
found that final variance in the DSS/USNOA coordinates 
to be very weakly dependent on the brightness of the
stars, and, for simplicity, we also assumed the uncertainty in the
CFH coordinates to be independent of the stellar flux, but
with positional variance smaller than that of the DSS positions 
by a factor of $200$.
The precision of our reference star positions should be limited
by the systematic error in the USNOA system, so
as future, more accurate, astrometric catalogs
become available it should be possible to improve
our astrometric solution.
Our solution corresponds essentially to a
maximum likelihood solution under the assumption 
that the position errors are
Gaussian distributed.
The resulting set of linear equations is of size
$N_{\rm exposures} \times N_{\rm chips} \times N_{\rm modes} + N_{\rm stars}$
which, with $N_{\rm modes} = 10$ for a cubic fit, is around 2000 and takes on the order of 1 hour to solve on
a low end workstation.

The hard work in solving this system of equations is in establishing the
labeling of stars by their $p$-index, so that we know that two stellar objects in
two different images are the same object.  To do this we made a sequence
of refined approximations for the mapping from pixel $x$-space to
sky coordinate $r$ in order to associate objects.  In this process
we made repeated use of `cross-correlation registration' which, given
a pair of catalogs containing a substantial number of objects in common, 
but with positions given in different coordinate systems,
automatically finds a scaling, rotation and translation which
maps one coordinate system into the other. The algorithm which
accomplishes this is described in
appendix \ref{section:acfregister}.
To get our first approximation to the mapping, accurate to maybe a few arc-seconds
at best, we generated approximate
detector plane coordinates $x_e$ for the brightest
few hundred objects in each of the exposures
by assuming the chips are simply laid out
on a regular grid with nominal spacing as described above, 
and found the scaling, rotation
and translation which maps $x_e$ coordinates to orthographic $r$ coordinates.
Note that ideally this step could have been avoided by using the
telescope pointing information encoded in the FITS image headers, but unfortunately
some of these turned out to be corrupted.
Armed with this first approximation we then extracted, 
for each image $i$,
a subset of the reference catalog lying under that image, and
then solved for a scaling, rotation and translation for each image
mapping pixel coordinate system $x_i$ onto
$r$ coordinates.
This is considerably more accurate than the
first solution as we are now approximating the mapping as a set
of `piecewise undistorted' patches rather than as a single large
undistorted patch. At this stage we also rejected saturated stars
from the CFH catalogs and also rejected some corrupted stellar
images by selecting on ellipticity.

Using this approximation for the $r$ coordinates of each object we
next
accumulated a 2K$\times$2K `object count' image covering the entire field
in $r$-coordinates,
and in which the pixel value
is the number of star detections. We then slightly smoothed
this image, ran our peak finder to generate a `master' catalog,
and to each of the objects assigned a unique
identifying index `$p$'.   We required a minimum of 4 separate detections
for an object to be included in this catalog to eliminate spurious detections.
We then merged each catalog in turn with this master
catalog and thereby inherited the $p$-values (by merging a pair 
of catalogs we mean finding pairs of detections whose positions coincide
to within some given tolerance).  Finally we concatenated these catalogs,
after rejecting extreme outliers,
and fed the result to a program which performs the least squares
solution of equation (\ref{eq:leastsquares}).  
This first solution is not perfect, as, due to the approximate
initial solution, a small fraction of the stars actually get detected as multiple
objects in the master catalog.  Armed with our first approximation
though, we can generate a refined object count
image and corresponding catalog, and
then obtain a refined least squares solution for the polynomial coefficients
$a_{ilm}$ and star positions.

\begin{figure}[htbp!]
\centerline{
\epsfig{file=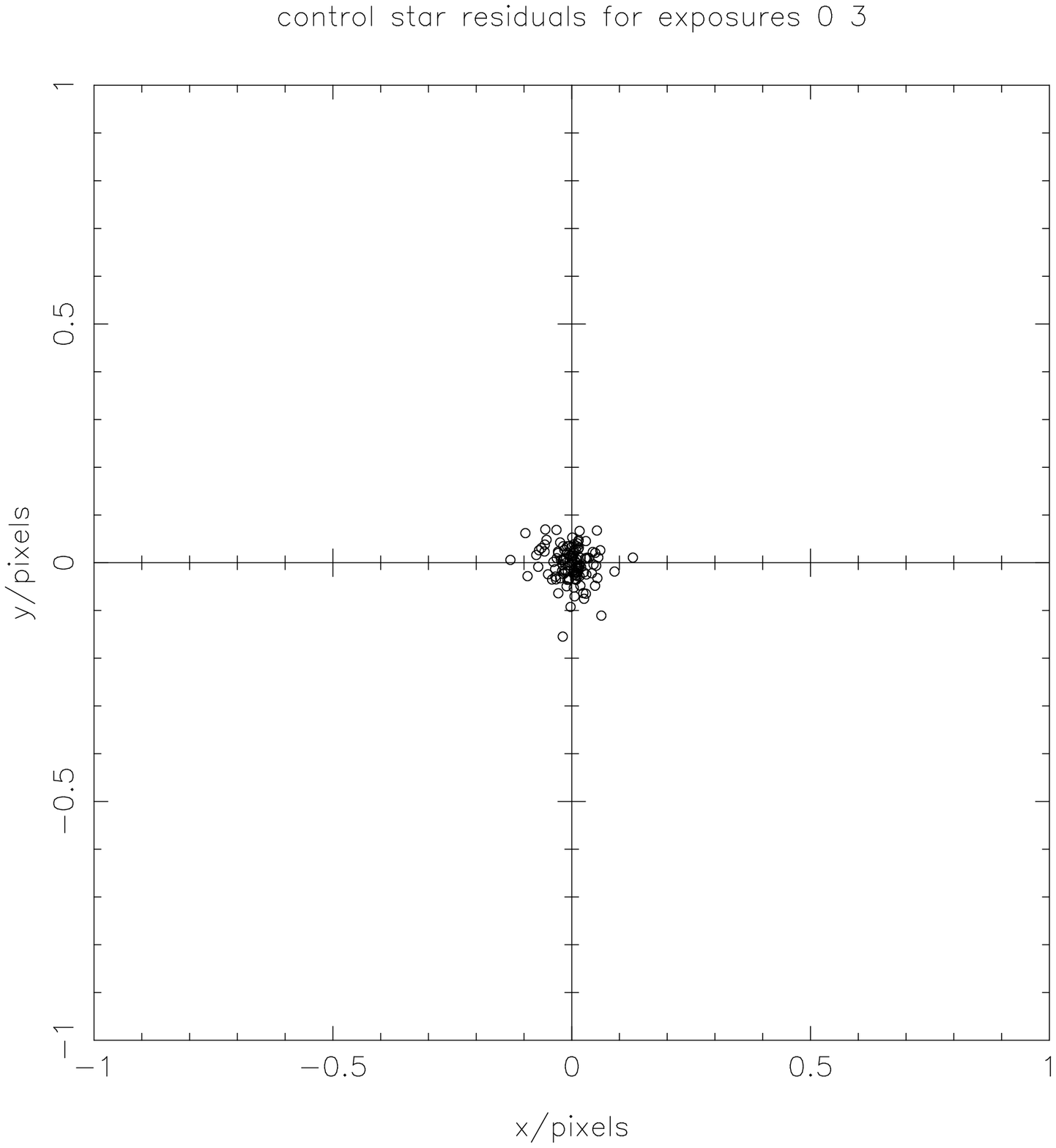,width=3.0truein}
\epsfig{file=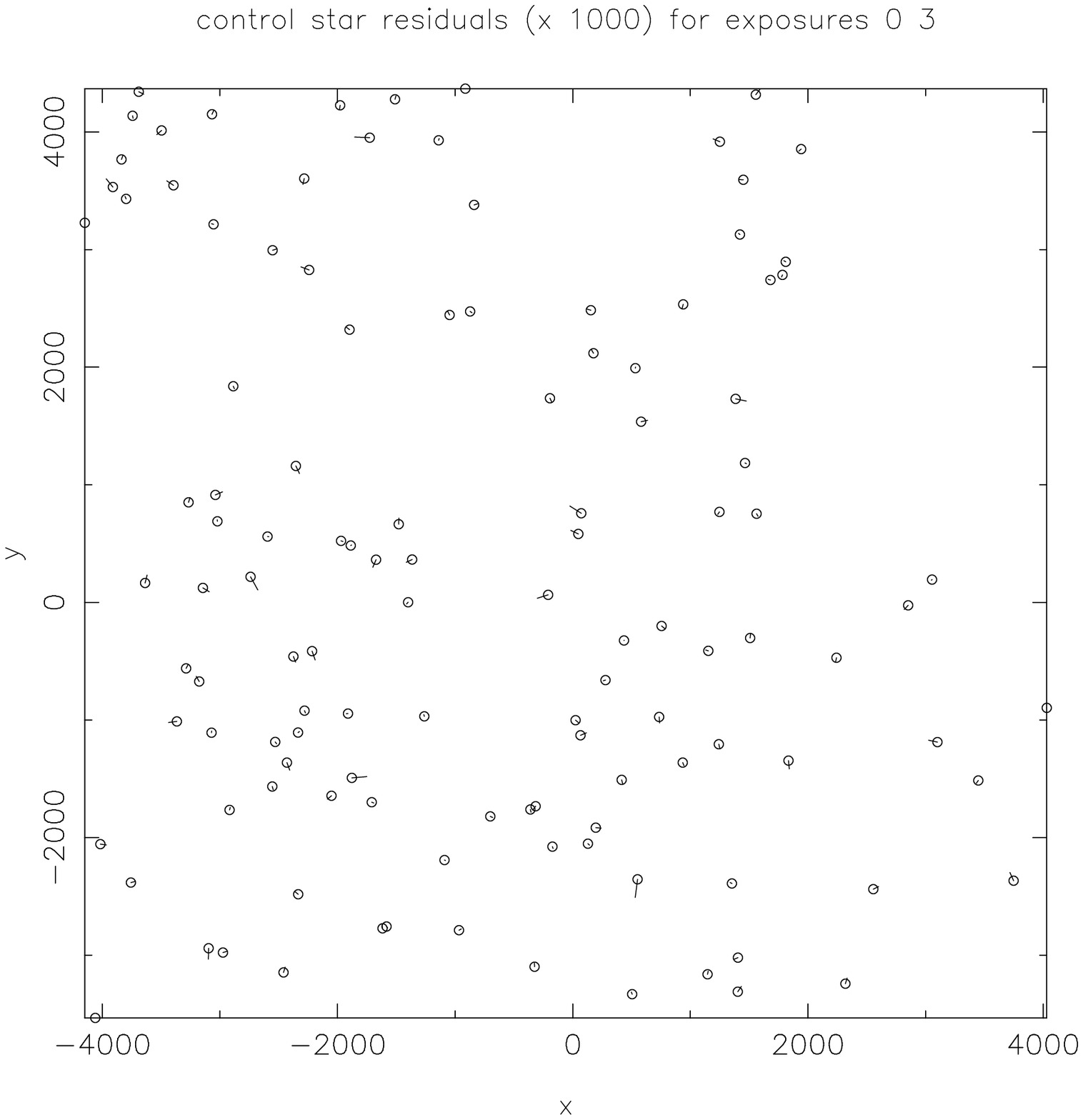,width=3.0truein}
}
\caption[Residuals in Astrometric Solution]{Residuals 
in the image mapping for the control sample. 
The left hand panel shows the differences between model $r$-values
for a set of control stars --- a randomly chosen
subset of $200$ stars which were not used in obtaining the
astrometric solution --- between a typical pair of exposures,
in this case the 0th and the 3rd.  The unit of length here is
the pixel size in the source images or $0''.207$.  The right hand
shows the same residuals, now  plotted as vectors with the base
of the vector placed at the object position and with the
length of the vector exaggerated by a factor 1000.  
These plots show that
the mapping is very accurate. The residual pairwise 
separation has rms of 7 milliarcsec
(per component), corresponding to a 1-particle positional uncertainty (smaller by
a factor $1/\sqrt{2}$) of about 5 milliarcsec.
The residuals appear to be approximately Gaussian distributed, and
we see no obvious systematic variation of the residuals over
the image; in particular, we find no tendency for unusually large residuals
along the `overlap' regions between the chips.}
\label{fig:residuals}
\end{figure}

As an objective check on how well this procedure worked, we
withheld a random subset of 20\% of the stars, and
did not use these in the registration solution.
After solving for the image mapping, we applied the solution  
to these stars and measured how
well their $r$-coordinates agreed.  
Typical results are shown in figure \ref{fig:residuals}.
This exercise gave a rms 
displacement of about $0''.007$, or about 1/30 of a pixel, for
the rms separation (one component) corresponding to a 1-particle
rms error of about $0''.005$.
With this degree of accuracy, any artificial shape distortion 
due to inaccuracy of the image mapping is, at worst,
on the order of $(\delta \theta / \theta_{\rm obj})^2$,
where $\delta \theta$ is typical error in the warping solution and
$\theta_{\rm obj}$ is the size of the objects,
and should be negligible.

The transformation 
coefficients $a_{ilm}$ we have thereby obtained give the mapping 
from $x_i$-space (being detector coordinates on the $i$th image) to
$r$-space (i.e.~it gives $r = r(x_i)$ as an explicit function of $x_i$).
For actually warping the images what is more useful is the inverse
mapping $x_i = x_i(r)$, since we need to compute, for each pixel in a target
image defined in $r$-space, what is the image of the pixel center in
$x_i$-space, so we need $x_i$ as an explicit function of $r$.  (Actually
one can perform the image warping using the forward transformation
but it turns out to be relatively expensive
in terms of computational effort). To obtain the inverse transformation
we generated a coarse regular grid of points which span the $2K\times 4K$ region
of $x_i$-space  occupied by the real pixels,  
applied the forward transformation to compute 
the model $r$-values and then fed these $x_i,r$ value pairs to
our least-squares program to
solve now for the inverse mapping
\begin{equation}
x_i = r + \sum\limits_{l,m} a'_{ilm} f_{lm}(r)
\end{equation}
to obtain the coefficients $a'_{ilm}$.
At this point we also set the final pixel scale in orthographic sky coordinates
onto which we will map the images.  
We adopted a pixel scale of $0''.15$ as compared to the
original pixel scale of about $0''.207$. 
This is desirable because the original
images are quite poorly sampled, the
stellar FWHM being $\simeq 0''.6$ or only about 3 pixels,
so aliasing sets in at relatively low spatial frequencies
where there is still substantial signal.  By interpolating and
resampling many images with random offsets
we suppress the aliaising to a large degree.
The PSF in the final image is then much better sampled, 
and quantities measured
either from stars or galaxies should be less sensitive to pixelization
effects. The main disadvantage is that the
storage space requirements are roughly doubled.

While the solution we have obtained maps the overlapping images onto
each other to impressive precision, the accuracy of the field distortion
we have derived leaves quite a lot to be desired.  
From the coefficients of the polynomial distortion model one can
readily compute the distortion tensor $\phi_{ij} = \d r_i / \d x_j$, the
shear $\gamma_\alpha = \half M_{\alpha l m} \phi_{lm}$, and the amplification as a function
of position on the final image.  
Away from the edges of the image these conform
quite nicely to the expected circular and approximately quadratic behavior, but 
close to the edges there are clear signs of errors in the solution.
Such errors are to be expected at some level due to the limited density
of stars and the relatively large uncertainty in the USNOA positions,
and particularly towards the edges of the field.
In fact what we see is somewhat larger than what we expect from simulations;
this may be due to systematic components to the USNOA positional error,
or to stars with anomalous position errors due to proper motions or
other effect.  This problem is exacerbated by the fact that the
chip in the SW corner of the array has a rather large region of
cosmetic defects on the side adjoining the rest of the array.
This has been masked out, rendering this chip almost disconnected from the
rest of the array.  For this chip, the errors in the distortion are very
large indeed ($\simeq 1-2\%$, as compared to the expected $\le 0.7\%$ shear
expected from the telescope itself). This is unfortunate, as it results
in an entirely spurious shear in the galaxies which we need to correct for.
We describe in \S\ref{subsec:getshapes} below how we have dealt with this.  

With more recent
observations (though of other targets) with the CFH12K camera
we have found that this problem can be avoided by taking
a sequence of 
preliminary short
2 minute exposures with large offsets (roughly half the chip
dimensions) and using these to obtain the astrometric solution.
From these short images one can generate a reference catalog to which one can
register the longer science exposures, these being taken with
relatively small offsets.
The solution obtained from the astrometric fields still suffers
from errors at the very edge, but these lie beyond the edge of the
region covered by the science exposures 
and therefore have little impact on the
final analysis.  A further advantage of the shorter exposures is a greater overlap
between non-saturated CFHT observed stars and stars in the USNOA
catalog. This further improves the quality of the astrometric solution,
and obviates the need for augmenting the USNOA catalog with DSS stars.

%% file: extinction.tex
\section{Correction for Extinction and Gain Variations}
\label{sec:extinction}

Using the catalogs of reference stars described in the 
previous section we solved, again in a least
squares manner, for a set of magnitude offsets (one set for the 
chips and one set for the exposures) which account for any 
variation in sensitivity between chips (and imprecision in the
zero-points determined from standard stars) and for varying
extinction between exposures.  These multiplicative 
corrections were very small;
typically $\sim 0.01$ magnitudes with maximum correction of $\simeq 0.04$
magnitudes, so we can conclude that the observing conditions were
accurately photometric.

%% file: psfmodel.tex
\section{Modeling the PSF}
\label{sec:psfmodeling}

The seeing in these images is very good, (FWHM $\simeq 0''.6$)
which means that
departures from the pure circular PSF expected from atmospheric 
turbulence become very noticeable.  These departures from circularity
in the PSF, which we shall denote by $g(x)$, most likely stem from several
sources, but principally from guiding errors 
and aberrations of the
telescope optics.  In the latter category there is a well known
astigmatism, thought to be caused by an imperfection in the primary 
mirror figure.  This is readily seen
as a variation in the PSF ellipticity as the
camera is moved up and down through focus. Consequently,  
this effect couples to any tilting of the chips
relative to the focal plane.  The result of this is a
a PSF which varies smoothly across any one chip, but
which jumps discontinuously as one passes to a neighboring chip.

\begin{figure}[htbp!]
\centering\epsfig{file=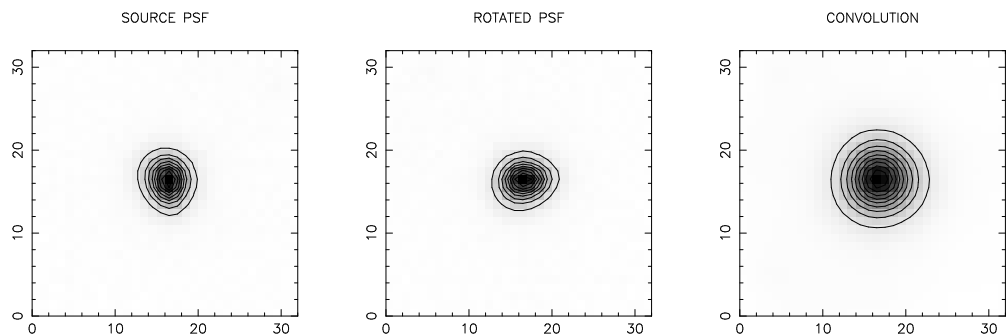,width=0.8 \linewidth}
\caption[Example PSFs]{Example PSFs.  The left hand panel shows a
typical PSF.  It is a realization of a 1st order polynomial fit to the
stars in the final image patch lying in the SE corner of the field. 
The middle panel shows the same PSF but rotated through 90 degrees
and the right hand panel shows the convolution of the two.}
\label{fig:psfexamples}
\end{figure}

This is a considerable nuisance for weak lensing studies
where one needs to accurately model and correct 
shapes of faint galaxies for variations in the PSF.
In dealing with single chip cameras one can model the PSF
as a smooth low-order polynomial.  
This is clearly inappropriate
here since the PSF variation in the summed images will have
step-like discontinuities. This is much harder to model accurately.
Another problem, though one which is not restricted to
mosaic cameras, is that great care must be taken in averaging
of images if stars in the final image are to be used to monitor
the PSF; the problem is that statistically robust averages like the
median will tend to ignore unusually distorted stellar images, while
the image of a faint galaxy on  the same image will not be
down-weighted.
The approach we have adopted here is
to re-convolve each of 
the source images with a kernel $g^\dagger = R_{\pi/2}(g)$,
a 90 degree rotated version of the PSF, in order to
render the final PSF approximately quadrupole free.
Figure \ref{fig:psfexamples} shows a typical PSF and
the result of convolving it with a $90^\circ$ rotated version.
Note that this reconvolution does not involve any actual
loss of information since it is applied after the photon
counting noise fluctuations have been realized in the measurement process.

For each source image we 
selected a sample of $\sim 100$ stars and
extracted a set of `postage stamp' images centered on the star and
32 pixels on a side.
To remove stars which were corrupted by cosmic rays etc we
computed a median of the stellar images, and for each star computed
the mean and maximum deviation from the median, and
rejected those with abnormally large deviations.
We then solved, by unweighted least squares, for a model
in which $g(x; x_0)$, this being the 
shape of a star lying at $x_0$ on the chip, is
a low order polynomial
\begin{equation}
g(x; x_0) = \sum\limits_{l,m} g_{lm}(x) f_{lm}(x_0)
\end{equation}
with mode functions $f_{lm}$
just as in equation (\ref{eq:flmdefinition}),
but now with image valued coefficients $g_{lm}(x)$. We used
a 1st order model, which seemed to adequately describe the PSF variations
we see.  
For each source image we then generated a `re-circularized'
image:
\begin{equation}
\label{eq:frecirc}
f_{\rm recirc}(x_0) =  \sum\limits_{l,m} f_{lm}(x_0) 
(R_{\pi/2}(g_{lm}) \otimes f)_{x_0}
\end{equation}

Using a 90-degree rotation is only an approximate method for
re-circularizing, but in this case works quite well, and
gives re-convolved images with PSF with quadrupole anisotropy
of $\lsim 1\%$, which is a great improvement over the
grossly anisotropic initial PSF.  Note that further low-level
anisotropy in the PSF in the summed images will result from
the image warping.  We describe in \S\ref{sec:photometry}
below how we have corrected for these effects.

%% file: warping.tex
\section{Image Warping and Averaging}
\label{sec:warping}

Armed with the results of the previous sections: the files containing 
the parameters $a'_{ilm}$ of the mapping from sky coordinates
$r$ to chip coordinates $x$ and the table of extinction/gain
corrections, we next generated a pair
of averaged images --- a raw and a re-circularized version.
As the final image spans $\sim 12,000$ pixels at our chosen $0''.15$
pixel scale, rather than generate a single image we chose to
generate a $6 \times 6$ grid of slightly overlapping
square patches each of size $2048 \times 2048$ pixels,
the central $2000 \times 2000$ sub-images of which form
a contiguous tiling of the sky.

More specifically, for each patch of the grid
and for each exposure, we inspected a set of
`bounding box' files generated in the registration process
to determine which source images contribute to that patch,
and generated a stack of $2048 \times 2048$ images, one per exposure, by applying the
polynomial warping transformations and also multiplicative
corrections for extinction etc.  
The warping was done with
bi-linear interpolation as described in appendix \ref{sec:2dinterpolation}.
The stack of images was then combined simply by taking the median.
The median is strictly less than optimal --- the
final variance is theoretically larger than a simple
averaging (assuming a large number of equally noisy source images) by a
factor $\pi/2$ \cite{ks77_1} --- but it is extremely robust to non-Gaussian
noise such as cosmic rays.  Moreover, we have found that in practice
the final variance is barely larger than that obtained from
more sophisticated methods involving rejection of outliers and
then averaging with weights proportional to $1 / \sigma_{\rm sky}^2$,
presumably because of low-level systematic errors in the sky
subtraction that start to become apparent when one averages large numbers
of images. It should be mentioned that in these data there were only
relatively minor fluctuations in the seeing from image to image, and
it may well be that the simple median averaging approach would not work well
when combining more heterogeneous data.
Also, a somewhat more sophisticated approach has been
used to combine the images in our UH8K blank field survey \cite{wkl99}.

The photometric scale of these final images is such that 
a single count in the averaged images corresponds to
a magnitude of $I = 32.39$, $V = 32.49$.  For our blank field
observations we also computed
an estimated inverse sky variance as the sum of the inverse variance
of the contributing images, as described below, and stored
$\sigma_{\rm sky}$ in an auxiliary image, but for MS0302 this was not done.

\subsection{Astrometric Information}
\label{subsec:astrometry}

The astrometric information which describes the mapping from pixel indices
in the averaged images
is stored in the FITS header for each
image patch using the convention of \citeNP{gc95}.
This `world coordinate system' is automatically
decoded by later versions of `saoimage' and `ximtool' so that the
cursor position read-out fields display $(\alpha, \delta)$ rather than
pixels. 

The claimed precision of the USNOA catalog is on the order
of $0''.25$, and so it is reasonable to assume that the final
uncertainty in our coordinates (which depends mainly
on the systematic component of the USNOA astrometry errors) is
no larger than this.
Small-scale random errors in our solution should be much smaller
than the systematic errors, and
our coordinates should be quite adequate for e.g.~the purpose
of making slit masks for multi-slit spectroscopy.

\subsection{Masking}
\label{sec:masking}

The saturated bright stars suffer from bleeding, reflection, and 
diffraction spikes. These confuse the faint object detection algorithm.  To
remove the resulting spurious detections
we have manually generated a set of mask
files, one for each patch of the quilt of the images, containing
rectangles
which enclose these trails and other obviously suspect parts of the final
averaged images. 

To do this we ran our object finding algorithm with a  low
significance threshold.  Displaying these objects superposed on the
average image proved to be a very effective way of identifying these
trails since
even very faint trails showed up as conspicuous streaks of
false detections and we were thereby able to interactively
generate the mask.

%% file: noise.tex
\section{Noise Properties and Limiting Magnitude}
\label{sec:noiseproperties}

These images are sky noise dominated; the number of photons being counted
being a Poisson distributed random variate with mean proportional to the
intensity, or, for the high counts we have here, Gaussian distributed to
a very good approximation. The faint galaxies are typically well below sky,
so the contribution to the noise associated with the background galaxy
signal is negligible and we can safely model the sky noise
in any source image as effectively flat and homogeneous.

We estimated the rms sky noise fluctuation $\sigma_{\rm sky}$ in the 
averaged images by measuring the curvature of the pixel value
distribution around the mode, as this is not greatly affected by the
skewing of the distribution by the signal.  There will be some
spatial variation in the sky variance due to the somewhat reduced
integration in the regions near the chip boundaries, but for simplicity
these have been ignored here.

The noise in the averaged images has short range correlations from the
mapping, interpolation and re-sampling.  
We have modeled the noise auto-correlation function using a simple
simulation in which incoherent white-noise images
were generated, interpolated and resampled 
just as for the real images.  Specifically, we dithered
these images on a $4\times 4$ grid of spacing $0.25$
source pixels.   This is not identical to the real data,
where the offsets were effectively randomly scattered
over the unit square, but as we have a reasonably large
number of images this simple model should fairly
accurately mimic the real noise correlations.
These images were then auto-correlated to obtain
a small FITS image of the noise ACF.
The ACF of the noise in the
re-circularized images can also then be obtained simply by convolving
this twice with the rotated PSF.

It is conventional to quote the limiting sensitivity as the
rms sky fluctuation expressed as a `magnitude per square arc-second', 
this being the rms fluctuation
per pixel if re-binned to $1''$ pixels and expressed as a magnitude.  
Allowing for the noise correlations we find that
in the well covered regions in these
summed images this is $I = 28.1$, $V = 28.7$.

%% file: objectdetection.tex
\section{Faint Object Detection}
\label{sec:objectdetection}

The faint object detection algorithm we have used is essentially
as described in \citeN{ksb95}, in that we smooth the images with a
sequence of `Mexican hat' filters of progressively larger size,
track the peak trajectories, and define an object to
be the peak of the significance (being the height of the peak
divided by the rms fluctuation for that smoothing scale) along the
trajectory.
The program which accomplishes this task is called {\tt hfindpeaks}.
The only major modification we have made is to allow for the
correlation of the noise in the images due to the interpolation and
re-binning.  This is done by having {\tt hfindpeaks} read
a small $32 \times 32$ pixel image of the noise ACF, generated
as described in \ref{sec:noiseproperties}, and then properly compute
the sky variance for each smoothing scale.

More specifically, we used a filter which is
a normalized Gaussian ball of scale $r_g$ minus another normalized concentric Gaussian of scale $2 r_g$.
The filter scale was varied from $0.5$ to $20.0$ with equally
spaced logarithmic intervals $\Delta r_g = 0.2 r_g$.
The algorithm was run on the raw (i.e.~not re-circularized) images, and
all objects with significance $\nu > 4$ were output (though
in the weak lensing analysis more conservative cuts were made).
The output of the program is a {\tt lc} format catalog (see appendix \ref{sec:catformat})
containing the items listed in table \ref{tab:hfindpeaks}.

\begin{table}[htbp!]
\caption{{\tt hfindpeaks} output}
\begin{tabular}{|c|c|l|}
\hline
name		&	type		&	description	\\
\hline
{\tt x}		&	2-vector	& peak location $x_i$	\\
{\tt lg}	&	scalar		& approximate flux $l_g$		\\
{\tt rg}	&	scalar		& Gaussian detection scale ($\times 0.66$) $r_g$	\\
{\tt eg}	&	2-vector	& smoothed peak ellipticity $e^g_\alpha$	\\
{\tt fs}	&	scalar		& smoothed peak height $f_s$	\\
{\tt nu}	&	scalar		& significance $\nu$		\\
\hline
\end{tabular}
\label{tab:hfindpeaks}
\end{table}

The quantity $l_g$ is a rather crude estimate of the
flux obtained assuming a Gaussian profile for the object in
question, and the
scaling of the detection radius is also based on a Gaussian shape.
The peak ellipticity $e^g_\alpha$ is computed from the
second derivatives of the smoothed image at the peak
location as $e^g_\alpha = M_{\alpha l m} \d_l \d_m f / \d_n \d_n f$.
The significance $\nu$ is the value of the smoothed field peak divided by the
rms of the noise fluctuations when smoothed at the same scale.

%% file: photometry.tex
\section{Faint Object Photometry}
\label{sec:photometry}

\subsection{Aperture Photometry}
\label{subsec:apphot}

The catalog generated by {\tt hfindpeaks} was then processed
by the command {\tt apphot} which performs basic aperture photometry.
The size of the aperture was set to be $3$ times the Gaussian
detection scale $r_g$ and from the pixels within this radius
we computed the quantities listed in table \ref{tab:apphot}.

\begin{table}[htbp!]
\caption{{\tt apphot} output}
\begin{tabular}{|c|c|l|}
\hline
name            &       type            &       description     \\
\hline
{\tt flux}         &       scalar        & sum of pixel values         \\
{\tt mag}        &       scalar          & magnitude           \\
{\tt rh}        &       scalar          & half light radius $r_h$      \\
{\tt rp}        &       scalar         	& Petrosian radius   $r_p$   \\
{\tt rql}        &       scalar          & radius containing 25\% of the light  \\
{\tt nqu}        &       scalar          & radius containing 75\% of the light \\
{\tt nbad}	&	scalar		& number of bad pixels \\
{\tt fmax}	&	scalar		& value of hottest pixel $f_{\rm max}$ \\
\hline
\end{tabular}
\label{tab:apphot}
\end{table}

\subsection{Shape Measurement}
\label{subsec:getshapes}

Following processing with {\tt apphot} we processed the catalog with the
command {\tt getshapes3} to obtain the weighted second moments and other
quantities that are used for weak shear analysis as described in \cite{kaiser99}.  
This added to the
catalog the entries listed in table \ref{tab:getshapes}.

\begin{table}[htbp!]
\caption{{\tt getshapes3} output}
\begin{tabular}{|c|c|l|}
\hline
name            &       type            &       description     \\
\hline
{\tt F}         &       scalar        & windowed flux  $F$       \\
{\tt q0}        &       scalar          & size       $q'_0$    \\
{\tt q}        &       2-vector          & polarization   $q'_\alpha$   \\
{\tt P0}        &       2-vector         	& size response $P'_0$    \\
{\tt P}        &       2x2 matrix          & polarization response $P'_{\alpha\beta}$ \\
{\tt R}	&	2-vector		& windowed flux response $R'_\alpha$ \\
{\tt Z}        &     2x2 matrix          & fourth moment $Z'_{\alpha\beta}$ \\
\hline
\end{tabular}
\label{tab:getshapes}
\end{table}

The windowed flux is defined in the continuum limit as
\begin{equation}
F = \int d^2 r\; w(r) f_s
\end{equation}
where $f_s$ is the re-convolved image,
and is approximated as a simple sum over pixels with no attempt
at sub-pixel precision.  The window function $w(r)$ was
taken to be a Gaussian with scale $\sigma$:
$w(r) = \exp(- r^2 / 2 \sigma^2)$.  
The second moments
\begin{equation}
q_A = (q_0, q_\alpha) = \half M_{A l m} \int
d^2 r \; w(r) r_l r_m f_s
\end{equation} 
are similarly approximated, but the quantity actually output
is rescaled by dividing by the
windowed flux: $q'_A = q_A / F$, so that the moments $q'_A$ have
dimension of (pixels)$^2$.  The quantities
$P_{A\beta}$, $R_\alpha$ are computed from the unreconvolved
image using a similarly discretized version of
equation (48) of \cite{kaiser99}, and again the
primes indicate that these
are also output after normalization by $F$. 

The computed polarization values $q'_\alpha$ here suffer from two
instrumental biases: residual anisotropy of the PSF and
PSF distortion introduced in applying the image mapping.
We now describe how we corrected for these.

\subsubsection{Correction for PSF Distortion from Image Mapping}

The problem here arose because, for simplicity, we applied the
recircularization process in detector coordinates, so the image
warping will have sheared the PSF and thus will 
necessarily have affected the shapes
of small objects.  This problem was exacerbated by the substantial
systematic error in our astrometric solution.
Had we instead contrived to generate a
re-convolved PSF $g^\dagger \otimes g$ that was
slightly anti-sheared so as to make the final PSF in the orthographic
sky projection circular, this problem would have been avoided and the \cite{kaiser99}
analysis could then be applied.
Alternatively, had we not taken out the telescope distortion (nor added
further erroneous distortion)
but measured the shapes in detector space then
again, the \cite{kaiser99} analysis could again be directly
applied to generate
a set of shear estimates $\hat \gamma_\alpha$, but with the
understanding that these would
in the final analysis need to be corrected for the telescope
distortion.  The latter presents no particular problem
since the telescope distortion is quite accurately measured from
our CFH12K astrometry observations on other fields.

The transformation from detector coordinates to
orthographic sky coordinates
is simply a shear applied after all
convolutions, so its effect on the moments $q_\alpha$
can be easily computed using the precepts of 
\citeN{ksb95}.  That work considered
the response of the ellipticity $e_\alpha \equiv
q_\alpha / q_0$ to a shear.  Here we have instead
moments normalized by the flux: $q'_\alpha = q_\alpha / F$,
but using the same line of reasoning, and specializing to the
case of a Gaussian window function, one finds that
applying the shear operator $f \rightarrow f' \simeq 
f - \gamma_\alpha M_{\alpha i j} r_i \d_j f$ gives a
`post-seeing shear response' of
\begin{equation}
\label{eq:shearresponse}
\Delta q'_\alpha = \gamma_\beta
\left[ 2 q'_0 \delta_{\alpha \beta} - Z'_{\alpha\beta} / 2 \sigma^2
+ 2 q'_\alpha q'_\beta / \sigma^2 \right]
\end{equation}
where
\begin{equation}
Z_{\alpha \beta} \equiv M_{\alpha i j} M_{\beta l m}
\int d^2 r \; w(r) r_i r_j r_l r_m f_s 
\end{equation}
and $Z'_{\alpha \beta} = Z_{\alpha \beta} / F$.
To apply this we generated an image of the shear
from the image mapping polynomial coefficients,
and the $q'_\alpha$ values were corrected by
subtracting (\ref{eq:shearresponse}). One should
really correct the last term here in (\ref{eq:shearresponse}),
which being
quadratic has a non zero noise induced expectation value,
but for simplicity, this rather small extra correction
was ignored.

\subsubsection{Correction from Residual PSF Anisotropy}

Re-convolving with a $90^\circ$ rotated PSF is only an
approximate re-circularizer.  The residual anisotropy
was on the order of 1 percent, and so can be important for small
galaxies.  We also found that low level PSF anisotropy over and above that
expected was introduced in the image mapping.  To correct for
these residual effects we have applied essentially
the KSB approximation: i.e.~we have assumed that the residual
anisotropy can be modeled as a convolution 
of a circular PSF with  a kernel $k(r)$
which is compact as compared to the overall width of the PSF.
Under this assumption, 
applying the `smear operator' $f \rightarrow f' \simeq
f + \half p_{lm} \d_l \d_m f$, where
$p_{lm} \equiv \int d^2 r \; r_l r_m k$ is the unweighted second
moment of the kernel, gives
a response
\begin{equation}
\label{eq:smearresponse}
\Delta q'_\alpha = {p_\beta \over \sigma^2}
\left[ (\sigma^2 - 2 q'_0) \delta_{\alpha \beta} 
+ Z'_{\alpha\beta} / 4 \sigma^2
- q'_\alpha q'_\beta / \sigma^2 \right]
\end{equation}
where $p_\alpha \equiv \half M_{\alpha l m} p_{lm}$.

To apply this correction we used (\ref{eq:smearresponse}) to
infer the $p_\beta$ values for a set of stars and fit these
as a 4th order spatial polynomial, and then corrected each
galaxy moment $q'_\alpha$ accordingly.

%% file: conclusions.tex
\section{Conclusions}

In this paper we have tried to describe in some detail 
the techniques we have developed for processing of images
taken with mosaic CCD cameras such as the UH8K and
the CFH12K.  These techniques have evolved continuously
over several years now, and are surely not yet optimal in
many regards, but some aspects of the analysis seem to work quite
well. We hope the description here may be helpful
either to people who wish to use our final images and
object catalogues --- which we hope to make available
electronically --- or to people who wish to use our
`imcat' software, which again is available on the internet.

It is a pleasure to acknowledge helpful conversations,
encouragement and also detailed comments on a draft
of this paper from Yannick Mellier, Thomas Erben, and
Emmanual Bertin.

%% file: acfregister.tex
\section{Cross-correlation Registration}
\label{section:acfregister}

The registration process described in \S \ref{sec:registration} is essentially
automatic, though there is some human intervention in the 
outlier rejection and refinement of the astrometric solution.
An essential ingredient is the algorithm for approximately
registering a pair of catalogs.  The algorithm only seeks to find
a scaling, rotation and displacement that effects the transformation
from one coordinate system to the other.  While crude, this may at least
allow one to merge or match the catalogs --- i.e.~find pairs whose
coordinates agree to some small tolerance --- and once the correspondence
between objects in the different catalogs is established one can
readily solve for more
elaborate transformation models (such as the low order spatial polynomial
models used extensively here).

\begin{figure}[htbp!]
\centering\epsfig{file=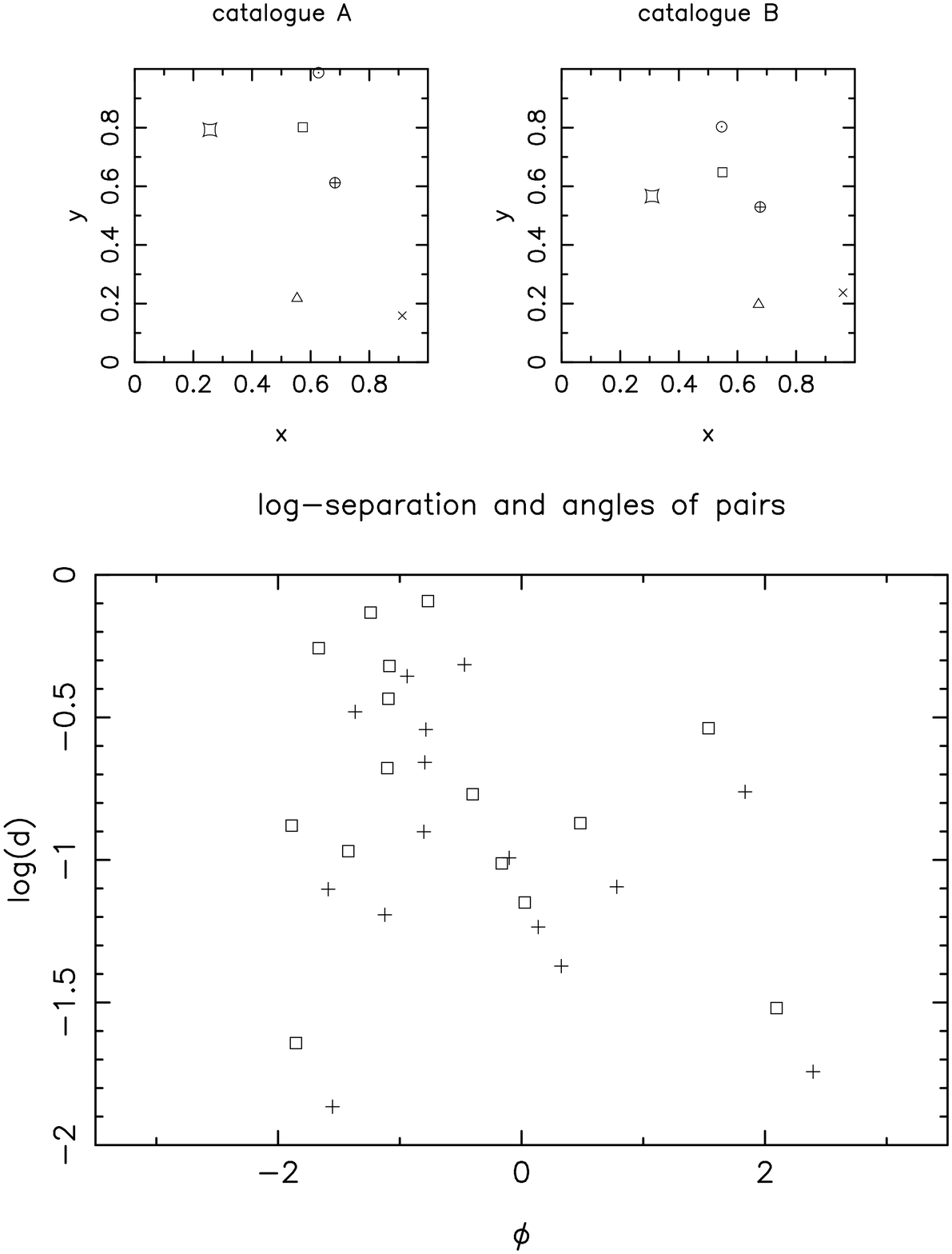,width=0.8 \linewidth}
\caption[Automatic Registration by Cross-Correlation]{Upper panels show the same set of 6 objects, but
with coordinates in catalog-B being scaled (by a factor 0.8),
rotated (by 0.3 radians) and shifted (by $(0.3, -0.1)$) with respect
to catalog-A. Lower panel shows a
plot of pairs from the each of the catalogs
in orientation {\it vs\/} log
separation space (with the pairs from catalog-B shaded).  It
can easily be shown, and is readily apparent from the figure, that
the pairs are simply shifted with respect to each other.}
\label{fig:acfregister}
\end{figure}

The problem is illustrated in the figure \ref{fig:acfregister}.
The solution we use is conceptually very simple; if for each catalog
we make a larger catalog containing 
all pairs of objects from the source catalog, 
and then plot these pairs
in $\phi$, $\log (d)$ space (where $\phi$ is the orientation angle of the
pair and $\log d$ is the natural logarithm of the pair separation) 
then the two pair catalogs should just be shifted with respect to each
other, with the shift in $\phi$ simply being the rotation between the
two coordinate systems, and the shift in $\log d$ being the 
logarithm of the scale factor. 

These rotation and scale factors can readily be determined by
autocorrelation using the FFT.  We simply generate a pair of images
of the counts of pairs in $(\phi, \log d)$ space and
compute their cross-correlation, which shows a strong peak at the
location of the real shift. 
Once we have the scale and rotation, we can apply these to
the first catalog.  This should now simply be a laterally shifted version of the
second catalog, and
we can solve for the shift by again making and cross-correlating 
a pair of images of the
counts of the objects.

Even if the input catalogs satisfy the scale, rotation and translation
transformation exactly, the result of {\tt acfregister} will be somewhat
imprecise due to the finite pixel size in the images used here.
Typically we use 
images of size $512 \times 512$ pixels, though the peak is located to 
fractional pixel precision, typically $1/30$ pixel or thereabouts,
giving a fractional precision in angle of $\sim 10^{-4}$.
The range of $\log d$ can be quite large, and if the image size were
set to encompass the whole range of $\log d$ then there 
would be loss of precision.  To avoid this problem 
we have wrapped the $\log(d)$ coordinate with a range of unity, 
so the precision in $\log(d)$ is of the same order as the angle.

This algorithm works quite well with real data, provided there is
a reasonable overlap between the objects in the two input catalogs,
and will still usually generate an acceptable solution
even if a substantial fraction of the
objects in one catalog are missing in the other.
We have found that the algorithm can become confused
with certain pairs of input catalogs --- because of the periodic boundary
conditions implied by the use of the FFT, it can, for instance, give a 
solution which is an alias of the desired solution (with the
shift incorrect by the side of the box say), but provided the
catalogs roughly cover the same area this is not a problem.

%% file: interpolation.tex
\section{Two-Dimensional Interpolation}
\label{sec:2dinterpolation}

Here we describe our sub-pixel interpolation scheme. We used interpolation
at two points in the process: first when we determined the centroids of
stars and second when we warped the images.  The interpolation schemes for
these two steps are different.

The object detection algorithm we used for detecting stars
for registration is very simple; we smoothed the image $f$ with
a Gaussian smoothing kernel with size similar to the psf and then
located the local maxima of the smoothed flux $F$
to obtain a integer pixel position $(i_x, i_y)$.
To refine this we computed first derivatives
\begin{equation}
\matrix{
F_x = (F_{i_x + 1, i_y} - F_{i_x - 1, i_y}) / 2 \cr
F_y = (F_{i_x, i_y + 1} - F_{i_x, i_y - 1}) / 2
}
\end{equation}
and second derivatives
\begin{equation}
\matrix{
F_{xx} = F_{i_x + 1, i_y} - 2 F_{i_x, i_y} + F_{i_x - 1, i_y} \cr
F_{yy} = F_{i_x, i_y + 1} - 2 F_{i_x, i_y} + F_{i_x, i_y - 1}
}
\end{equation}
and then computed refined positions according to
\begin{equation}
\matrix{
x = i_x - F_x / F_{xx} \cr
y = i_y - F_y / F_{yy}
}
\end{equation}
Our pixel coordinate labeling convention differs from the
FITS standard in which pixel centers have integer values with
FORTRAN style unit offset indexing --- so the physical region covered by a
$N\times N$ chip is defined to be $0.5 < x,y < N + 0.5$.  Here
our pixel centers have half-integer values and we use the
`c'-style zero offset indexing so we have $0 < x,y < N$, and
the center of the corner pixel, for instance, is $(x,y) = (0.5,0.5)$.

When we warped the images we used a different type of interpolation.
The pixel values are samples of a continuous
function $f$ (being the convolution of the sky flux with the
box-like pixel response function)
on a grid points $(i_x + 1/2, i_y + 1/2)$.  A point $(x,y)$
lies within a square defined by four of these samples.  To interpolate
the $f$ value we added a fifth sample at the center of the square
which is the average of the four corner values.  Joining the four corners to
the center divides the square into four triangles, with the interpolation
point lying in one of these.  We took as our estimate of 
$f(x,y)$ the height at $(x,y)$ of the plane which passes through the three
samples at the corners of this triangle.
The interpolated $f$ values thus
obtained are continuous, but have discontinuity of gradient along
the vertical, horizontal and diagonal
lines connecting the original sample grid.

%% file: lcformat.tex
\section{Catalog Format}
\label{sec:catformat}

Along with the summed images, our database, which
we plan to make available electronically, also contains catalogs
of objects and numerous auxiliary tabular information.  
The format of these catalogs and other tables is defined by
the program `{\tt lc}' (for {\underline l}ist {\underline c}atalog)
which, as its name implies, in its default mode, simply generates
an ASCII format listing of a catalog.  However, with its numerous command
line options, {\tt lc} becomes a fairly versatile filter for
manipulating catalogs.

The {\tt lc} program is very similar to the UNIX command {\tt awk}, in that it
processes catalogs one object at a time; reading an object from standard 
input - 
performing some manipulation on the contents of the object as
specified by instructions supplied on the
command line - and writing the result to standard output,
but with the
following distinctions: a) fields or entries in the object are
referred to by symbolic names rather than by column number; b) entries may
be scalars, vectors or matrices of arbitrary rank 
(there is also some limited support for textual entries) and c) {\tt lc}
can read and write in binary format, resulting in a large gain
in efficiency.

Operations to be performed are specified as command line strings
in a simple `reverse-Polish' notation.
All of the standard {\tt c} math library functions and operators
as well as a number of specialized functions and operators (such as
vector products, matrix inversion etc.) are supported.

Our photometry packages are implemented as `filters' which read
{\tt lc}-format catalogs and add size, magnitude, shape etc information
consecutively.  All the auxiliary files used in the reduction
process described here are stored as {\tt lc}-format catalogs for ease
of manipulation.
Users of our database may find {\tt lc} useful for extracting
variables of interest from our catalogs and for applying selection
criteria to select sub-catalogs.